\documentclass[a4paper,12pt]{article}
\pdfoutput=1
\usepackage{jheppub}

\title{Refined Predictions for Starobinsky Inflation and Post-inflationary Constraints\\ in Light of ACT}
\author[a]{Manuel Drees}
\author[b]{and Yong Xu}

\emailAdd{drees@th.physik.uni-bonn.de}
\emailAdd{yonxu@uni-mainz.de}
\affiliation[a]{\it Bethe Center for Theoretical Physics and Physikalisches Institut, Universit\"at Bonn,\\Nussallee~12, 53115 Bonn, Germany}
\affiliation[b]{{\it PRISMA$^+$} Cluster of Excellence and Mainz Institute for Theoretical Physics\\
	Johannes Gutenberg University, 55099 Mainz, Germany}
\abstract{Recent measurements from the Atacama Cosmology Telescope
	(ACT), combined with Planck and DESI data, suggest a higher value
	for the spectral index $n_s$. This places Starobinsky inflation at
	the edge of the $2\sigma$ constraints for a number of e-folds
	$N_\star$ around $60$ when using the usual analytical
	approximations. We present refined predictions for Starobinsky
	inflation that go beyond the commonly used analytical
	approximations. By evaluating the model with these improved
	expressions, we show that for $N_\star \gtrsim 58$ it remains
	consistent with current observational constraints at the $2\sigma$
	level. Additionally, we examine the implications of the ACT results
	for post-inflationary reheating parameters. Specifically, we find a
	lower bound on the effective equation of state parameter during
	reheating of approximately $\omega \gtrsim 0.462$; this excludes
	purely perturbative reheating, which leads to $\omega \simeq 0$. We
	also show that the reheating temperature is constrained to be
	$T_{\text{rh}} \lesssim 2 \times 10^{10}~\text{GeV}$, assuming
	$\omega \leq 1$. Furthermore, we find that the predictions for the
	spectral index and tensor-to-scalar ratio can lie within $1\sigma$
	of the recent ACT constraints if the reheating temperature satisfies
	$4~\text{MeV} \lesssim T_{\text{rh}} \lesssim 10~\text{GeV}$ for
	$0.8 \lesssim \omega \leq 1$.}

\begin{document}
	\begin{flushright}
		MITP-25-033
	\end{flushright}
	\maketitle

	\section{Introduction}
	Cosmic inflation offers an elegant solution to several key problems in
	cosmology~\cite{Starobinsky:1980te, Guth:1980zm, Linde:1981mu,
		Albrecht:1982wi}, with detailed reviews available in
	Refs.~\cite{Olive:1989nu, Martin:2013tda, Ellis:2023wic}. It predicts
	a nearly scale-invariant power spectrum, which is typically
	characterized by the spectral index $n_s$. The Planck 2018 results
	report a value of $n_s = 0.965 \pm 0.004$ \cite{Planck:2018vyg,
		Planck:2018jri}. More recent data from the Atacama Cosmology
	Telescope (ACT) suggest a higher value for $n_s$ \cite{ACT:2025fju,
		ACT:2025tim}. Specifically, Ref.~\cite{ACT:2025fju} shows that
	combining Planck, ACT, and DESI data~\cite{DESI:2024uvr, DESI:2024mwx}
	yields $n_s = 0.974 \pm 0.003$. Additionally, Ref.~\cite{ACT:2025tim}
	finds that Starobinsky inflation~\cite{Starobinsky:1980te} is located
	at the $2\sigma$ boundary of the combined constraints, with the number
	of e-folds before the end of inflation estimated at around $60$. These
	results have sparked several discussions and (or) alternative
	proposals in response to the ACT results ~\cite{Kallosh:2025rni,
		Aoki:2025wld, Berera:2025vsu, Dioguardi:2025vci, Salvio:2025izr,
		Dioguardi:2025mpp, Rehman:2025fja, Gao:2025onc, He:2025bli,
		Gialamas:2025kef}.
	
	In this work, we revisit the predictions within the Starobinsky
	inflationary framework. We begin by recalling that, in the limit of
	large $N_\star$ (the number of e-folds before the end of inflation),
	the inflationary predictions of the Starobinsky model can be
	approximated as \cite{Ellis:2013nxa, Kallosh:2013maa, Kallosh:2013yoa,
		Motohashi:2014tra}
	\begin{align}\label{eq:ns_r_approximation}
		n_s &\simeq 1 - \frac{2}{N_\star}, \quad r \simeq \frac{12}{N_\star^2}\,, 
	\end{align} 
	where $n_s$ is the spectral index and $r$ is the tensor-to-scalar ratio.
	
	Given the recent advances in the measurement of $n_s$, it has become
	essential to employ more accurate theoretical predictions when
	confronting observational results. Motivated by this, one of the main
	objectives of this work is to provide simple yet more precise
	expressions for the inflationary observables, particularly for $n_s$,
	within the Starobinsky model. Additionally, we aim to investigate the
	resulting constraints on reheating parameters derived from the recent
	combined measurements of $n_s$; this is another goal of this work.
	
	The rest of this article is organized as follows. In
	Section~\ref{sec:Starobinsky}, we revisit Starobinsky inflation,
	focusing on the derivation of more precise expressions for its
	inflationary predictions. In Section~\ref{sec:comparison}, we compare
	these results with the commonly used approximations and confront them
	with the recent ACT results. We then investigate the implications for
	post-inflationary reheating parameters in Section
	\ref{sec:Reheating_constraints}. Finally, we summarize the main
	findings of this work in Section~\ref{sec:conclusion}.
	
	\section{Starobinsky Inflation and Predictions}
	\label{sec:Starobinsky}
	
	In this section, we briefly revisit Starobinsky
	inflation~\cite{Starobinsky:1980te}, one of the earliest predictive
	models of inflation. The theory is based on a modification of general
	relativity, and is described by the action
	\begin{align}\label{eq:action}
		S \supset \frac{M_{\rm Pl}^2}{2} \int d^4x \sqrt{-g}
		\left( R + \frac{1}{6M^2} R^2 \right),
	\end{align}
	where $g$ is the determinant of the metric $g_{\mu \nu}$, $R$ is the
	Ricci scalar, $M_{\rm Pl}$ denotes the reduced Planck mass, and $M$ is
	a mass scale associated with the inflationary dynamics. The presence
	of the $R^2$ term leads to a period of accelerated expansion in the
	early Universe, without invoking an explicit scalar inflaton field.
	
	By performing a conformal transformation to the Einstein frame, the
	theory can be recast as a scalar field theory with a canonical kinetic
	term and an effective potential given by \cite{Ellis:2013nxa,
		Kallosh:2013yoa, Motohashi:2014tra, Ellis:2021kad}
	\begin{align}\label{eq:V}
		V(\phi) = \frac{3 M^2 M_{\rm Pl}^2}{4} \left(1 - e^{-\sqrt{\frac{2}{3}}\frac{\phi}{M_{\rm Pl}}} \right)^2.
	\end{align}
	This potential features a plateau at large inflaton field values,
	supporting slow-roll (SR) inflation.
	
	To obtain the inflationary predictions, we apply the SR formalism and
	define the following SR parameters
	\begin{equation}
		\epsilon_V \equiv \frac{M_{\rm Pl}^2}{2} \left(\frac{V^{\prime}}{V}\right)^2,
		\quad
		\eta_V \equiv M_{\rm Pl}^2\, \frac{V^{\prime \prime}}{V}  \,, 
	\end{equation}
	whose magnitude must be smaller than one during the accelerated
	expansion of the Universe. Here, $\prime$ means the derivative of the
	potential with respect to $\phi$. Inflation ends when the field
	reaches the value $\phi = \phi_{\text{end}}$ defined as
	$\epsilon_V (\phi_{\text{end}}) = 1$ in the current large field setup.
	The total number of e-folds $N_\star$ between the time when the Cosmic
	Microwave Background (CMB) pivot scale $k_{\star} = 0.05$~Mpc$^{-1}$
	first crossed the horizon (with a field value $\phi = \phi_\star$)
	until the end of inflation is given by
	\begin{align}\label{eq:N_star}
		N_\star &= \int^{\phi_\star}_{\phi_{\text{end}}} \frac{1}
		{\sqrt{2\, \epsilon_V(\phi)}}\, \frac{d\phi}{M_{\rm Pl}} \,.
	\end{align}
	The power spectrum of the curvature perturbation $A_s$, the spectral
	index $n_s$, and the tensor-to-scalar ratio $r$ are respectively given
	by
	\begin{align}
		A_s &= \frac{1}{24\pi^2\, \epsilon_V}\, \frac{V}{M_{\rm Pl}^4}\,,\\
		n_s &= 1- 6\epsilon_V + 2\eta_V\,, \label{eq:ns_def}\\
		r &=16 \epsilon_V \,,
	\end{align}
	in the SR formalism.\footnote{The ACT result \cite{ACT:2025fju} $n_s = 0.974 \pm 0.003$
		assumes constant $n_s$. This could be cause for concern, since ACT and
		Planck (or the BAO) probe rather different scales. In fact, ACT \cite{ACT:2025tim} also finds weak evidence for positive running $\alpha$ of $n_s$, i.e. $n_s$ becoming
		larger at larger wave vectors or smaller length scales. However, the Starobinsky model predicts $\alpha \simeq -(1-n_s)^2/2$ to be small and
		negative. Including the running of $\alpha$ will therefore not improve the agreement between this model and ACT data.} The central value of the power spectrum is
	$A_s = 2.1 \times 10^{-9}$ \cite{Planck:2018vyg}, which allows one to
	determine the scale of $M$.\footnote{In practice, the parameter $M$
		can be expressed as
		$M = \pi\, M_{\rm Pl} \sqrt{\frac{A_s}{6}} \left(1 + 2\sqrt{4 -
			3\,n_s} - 3\,n_s\right)$, which evaluates to approximately
		$10^{-5}\, M_{\rm Pl}$ for $n_s \simeq 0.97$.} We note that the
	inflationary predictions $n_s$ and $r$ are independent of $M$, as will
	be shown below.
	
	Using the potential in Eq.~\eqref{eq:V}, we find the tensor-to-scalar
	ratio and spectral index can be written as
	\begin{align} \label{eq:r}
		r& =\frac{64}{3\left( e^{\sqrt{\frac{2}{3}} \frac{\phi_\star}{M_{\rm Pl}}}
			- 1 \right)^2}  \,,
	\end{align}
	\begin{align} \label{eq:ns}
		n_s&=\frac{3 \left( e^{\sqrt{\frac{2}{3}} \frac{\phi_\star}{M_{\rm Pl}}}
			- 1 \right)^2 - 8
			\left(e^{\sqrt{\frac{2}{3}} \frac{\phi_\star}{M_{\rm Pl}}} + 1\right) }
		{3 \left( e^{\sqrt{\frac{2}{3}} \frac{\phi_\star}{M_{\rm Pl}}}-1\right)^2}\,.
	\end{align}
	Note that in the large-field limit, where the potential becomes flat,
	one finds the expected behavior: $r \to 0$ and $n_s \to 1$.
	
	The field value at the end of inflation is given by
	\begin{align}\label{eq:phiend}
		\phi_{\text{end}} = M_{\rm Pl} \sqrt{\frac{3}{2}} \ln\left( 1
		+ \frac{2}{\sqrt{3}} \right)\,.
	\end{align}
	From Eq.~\eqref{eq:ns}, one can solve for $\phi_\star$ with a given $n_s$:
	\begin{align} 
		\phi_\star = M_{\rm Pl}\sqrt{\frac{3}{2}} \ln
		\left[\frac{7+ 4\sqrt{4 -3\,n_s} -3\, n_s}{3(1-n_s)}\right]\,,
	\end{align}
	which, with  Eq.~\eqref{eq:r}, yields
	\begin{align}\label{eq:r2}
		r
		& = \frac{4}{3} \left[3(1-n_s)+2- 2\sqrt{3(1-n_s) +1} \right]\,.
	\end{align}
	We note that Eq.\eqref{eq:r2} has been presented earlier in
	Ref. \cite{Garcia:2023tkk}.
	
	The total number e-folds $N_\star$ in Eq.~\eqref{eq:N_star} can also
	expressed as a function of $n_s$:
	\begin{align}\label{eq:N_star2}
		N_\star & = \int^{\phi_\star}_{\phi_{\text{end}}} \frac{1}{\sqrt{2 \epsilon_V}}
		\frac{d\phi}{M_{\rm Pl}}
		= \frac{3}{4} \left[ e^{\sqrt{\frac{2}{3} }\frac{\phi_\star}{M_{\rm Pl}}}
		- e^{\sqrt{\frac{2}{3} }\frac{\phi_{\text {end}}}{M_{\rm Pl}}}
		- \sqrt{\frac{2}{3}} \frac{\left(\phi_\star- \phi_{\text{end}}
			\right)}{M_{\rm Pl}} \right]\,\nonumber \\
		&=
		\frac{2-\sqrt{3} (1-n_s) + 2\sqrt{3(1-n_s)+1} }{2(1-n_s)} -\frac{3}{4}\ln \left[
		\frac{4+3(1-n_s)+4 \sqrt{3(1-n_s)+1}}{(1-n_s)(2 \sqrt{3} +3 ) }
		\right]\,\nonumber \\
		& = \frac{2-\sqrt{3} t +2\sqrt{3t+1} }{2t} -\frac{3}{4}\ln \left[
		\frac{4+3t+4 \sqrt{3t+1}}{t(2 \sqrt{3} +3 ) } \right]\,,
	\end{align}
	where we have introduced a small parameter $t \equiv 1 - n_s$ in the
	last step.
	
	We now discuss how the exact results obtained above reduce to the
	commonly used approximations shown in
	Eq.~\eqref{eq:ns_r_approximation}. Noting that $n_s$ is close to
	unity, we write the last line of Eq.~\eqref{eq:N_star2} as
	\begin{align}\label{eq:N_star3}
		N_\star &\simeq \frac{2}{t} + \frac{1}{4} \left[6 - 2\sqrt{3}
		+ 3\ln\left(\frac{3 + 2\sqrt{3}}{8}\right) + 3\ln t \right]
		+ \mathcal{O}(t)\,\nonumber \\
		& \simeq \frac{2}{1-n_s} +\frac{3}{4}\ln (1-n_s) +0.47  +\mathcal{O}(1-n_s)\,.
	\end{align}
	Note that the second term in Eq.~\eqref{eq:N_star3} is negative and of
	order unity, which effectively reduces the value of $N_\star$ for
	given $n_s$. Similarly, the tensor-to-scalar ratio given in
	Eq.~\eqref{eq:r2} can be expanded as
	\begin{align}\label{eq:r3}
		r & \simeq 3t^2 + \frac{9}{2}t^3 + \mathcal{O}(t^4)\,\nonumber \\
		& \simeq  3(1-n_s)^2 + \frac{9}{2}(1-n_s)^3 + \mathcal{O}[(1-n_s)^4]\,.
	\end{align}
	We note that the parameter $t$, or equivalently
	$1 - n_s$, is small; in particular, $t = 6\epsilon_V - 2\eta_V$, as
	given in Eq.~\eqref{eq:ns_def}. In the following, we will neglect
	higher-order terms in $t$ in Eq.~\eqref{eq:N_star3} and
	Eq.~\eqref{eq:r3}, which are much smaller than the current observational uncertainty on $1-n_s$. 
	
	It is interesting to observe that Eq.~\eqref{eq:N_star3} can be solved
	iteratively to obtain a simple yet precise relation between $n_s$ and
	$N_\star$; combined with the first term in Eq.~\eqref{eq:r3}, this
	yields
	\begin{align}\label{eq:ns_iteration}
		n_s \simeq 1- \frac{2}{N_\star- \frac{3}{4}\ln \left(\frac{2}{N_\star}
			\right)}\,,
		\quad r \simeq \frac{12}{\left[N_\star- \frac{3}{4}
			\ln \left(\frac{2}{N_\star}\right)\right]^2}\,, 
	\end{align} 
	which match very well with the exact results shown in
	Eq.~\eqref{eq:r2} and Eq.~\eqref{eq:N_star2}.  To the best of our
	knowledge, the expression for $n_s$ presented in
	Eq.~\eqref{eq:ns_iteration} has not previously been shown in the
	literature. We consider it to be one of the main results in this work,
	and we advocate its use when confronting the Starobinsky model with
	experimental data.
	
	We now discuss how to reproduce the usual approximation
	Eq.~\eqref{eq:ns_r_approximation} using
	Eq.~\eqref{eq:ns_iteration}. By retaining only the first terms both in
	Eq.~\eqref{eq:N_star3} and Eq.~\eqref{eq:r3} or dropping the term
	$-\frac{3}{4}\ln \left(\frac{2}{N_\star}\right)$ in
	Eq.~\eqref{eq:ns_iteration}\footnote{Note that the value of
		$-\frac{3}{4}\ln\left(\frac{2}{N_\star}\right)$ is approximately
		$2.41$ for $N_\star = 50$ and $2.55$ for $N_\star = 60$, which can
		lead to a noticeable shift in $n_s$.}, we recover the commonly used
	approximate expressions in Eq.~\eqref{eq:ns_r_approximation}:
	\begin{align}\label{eq:approximation}
		n_s \simeq 1 - \frac{2}{N_\star}, \quad r \simeq \frac{12}{N_\star^2}\,.
	\end{align}
	Eq.~\eqref{eq:approximation} implies a relation
	\begin{align}\label{eq:approximation2}
		r \simeq 3 (1 - n_s)^2\,,
	\end{align}
	which corresponds to the first term of Eq.~\eqref{eq:r3}. Since there
	is no term $\propto \ln(1-n_s)$ in Eq.~\eqref{eq:r3},
	Eq.~\eqref{eq:approximation2} also holds for our refined approximation
	Eq.~\eqref{eq:ns_iteration}, up to corrections of relative order
	$1-n_s$
	
	Before closing this section, we note that the spectral index $n_s$ in
	the approximation Eq.~\eqref{eq:approximation} is {\it underestimated}
	for a given number of e-folds $N_\star$, as is evident when compared
	with Eq.~\eqref{eq:ns_iteration}. These comparisons will be made more
	explicit in the next section, where we also discuss the implications
	with current experimental constraints.
	
	\section{Comparison between Exact and Approximate Results}
	\label{sec:comparison}
	\begin{figure}[!ht]
		\def\sepf{0.8}
		\centering
		\includegraphics[scale=\sepf]{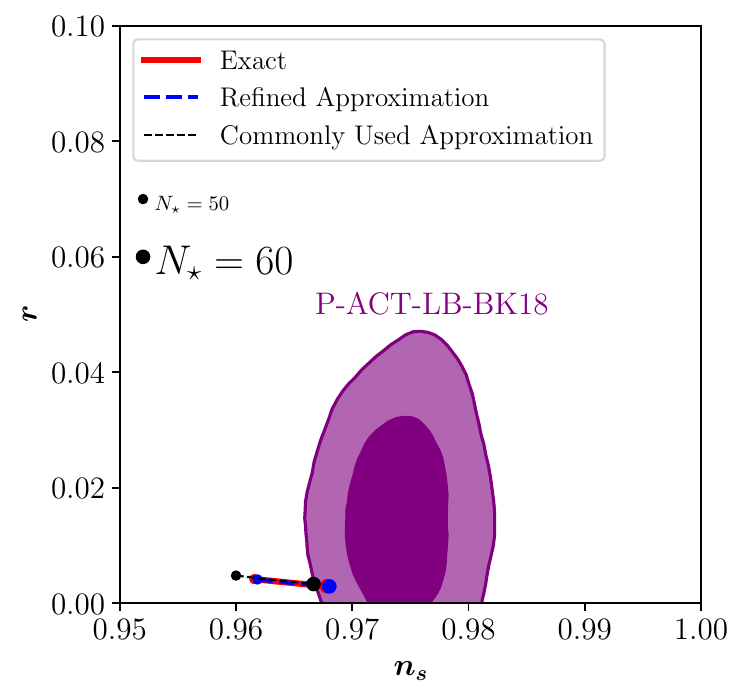}
		\caption {Comparison of the exact expression in Eq.~\eqref{eq:r2} with
			the refined approximation in Eq.~\eqref{eq:ns_iteration} and the
			commonly used approximation in Eq.~\eqref{eq:approximation} for $r$
			and $n_s$ within the framework of Starobinsky inflation.}
		\label{fig:ns_r}
	\end{figure} 
	
	In this section, we compare the approximations in
	Eqs.~\eqref{eq:approximation} with the exact result given in
	Eq.~\eqref{eq:r2}, as well as with the refined approximation in
	Eq.~\eqref{eq:ns_iteration}, and confront them with current
	observational data.
	
	The results are presented in Fig.~\ref{fig:ns_r}, where we show the
	predictions in the $r$–$n_s$ plane, overlaid with the $2\sigma$
	combined constraints from Planck, ACT, DESI, and BICEP/Keck
	data~\cite{BICEP:2021xfz} (P-ACT-LB-BK18), shown in purple and
	reproduced from Fig.~10 of Ref.~\cite{ACT:2025tim}. The red solid line
	shows the exact result based on Eq.~\eqref{eq:r2}, where $n_s$ is
	determined by Eq.~\eqref{eq:N_star2} for a given number of
	e-folds. The blue dashed line depicts the refined approximation using
	Eq.~\eqref{eq:ns_iteration}. The black dashed line corresponds to the
	approximate relation obtained using Eqs.~\eqref{eq:approximation} and
	\eqref{eq:approximation2}. The small and big dots denote
	$N_\star = 50$ and $N_\star = 60$, respectively.
	
	As illustrated, the commonly used approximation in
	Eq.~\eqref{eq:approximation} (black dashed) suggests that Starobinsky
	inflation with $N_\star \leq 60$ lies near the boundary of the
	$2\sigma$ allowed region. In contrast, we show that the exact
	prediction (red solid), based on Eq.~\eqref{eq:r2}, remains within the
	$2\sigma$ allowed region at $N_\star=60$. It is also evident that the
	approximation in Eq.~\eqref{eq:approximation} {\it underestimates} the
	spectral index $n_s$. Furthermore, we demonstrate that the refined
	approximation Eq.~\eqref{eq:ns_iteration} (blue dashed) provides a
	better fit to the exact prediction for $n_s$. These comparisons
	underscore the importance of employing more accurate theoretical
	expressions when interpreting current and future cosmological data.
	
	Before closing this section, we note that we have so far considered
	$N_\star \in [50, 60]$, which is typically a good
	assumption~\cite{Liddle:2003as, Dodelson:2003vq}. However, $N_\star$
	can also be larger, depending on the specifics of the
	post-inflationary expansion history~\cite{Liddle:2003as,
		Tanin:2020qjw}. As we will demonstrate in the next section, a larger
	$N_\star$ can arise when matching with a post-inflationary phase that
	has a stiff background. As can be seen in Fig.~\ref{fig:ns_r}, the
	prediction for $n_s$ can lie well within the $2\sigma$ region (even
	$1\sigma$ ) if $N_\star$ is sufficiently large. This underscores the
	necessity of consistently matching inflationary predictions with the
	subsequent expansion history, a topic we will discuss in the next
	section.
	
	\section{Constraints on Reheating}
	\label{sec:Reheating_constraints}
	The connection between the dynamics of reheating and predictions for
	$n_s$ has also been made in Ref.~\cite{Mishra:2021wkm}, in the context
	of specific models including the Starobinsky model we are analyzing
	here. In this section, we investigate the implications of the recent
	ACT results for post-inflationary reheating parameters. Inflation must
	eventually come to an end, allowing the transfer of energy from the
	inflaton field to other particles, whose subsequent interactions lead
	to the formation of a thermal bath of Standard Model (SM)
	particles. The detailed dynamics of reheating, however, remain unknown
	and can be highly complex; see Refs.~\cite{Allahverdi:2010xz,
		Amin:2014eta, Lozanov:2019jxc} for reviews.
	
	To facilitate reheating, the current setup shall be extended to allow
	the inflaton to decay or annihilate. Indeed, after performing the conformal transformation, couplings between the inflaton and matter fields naturally arise.  Here, we remain
	agnostic about the specifics of reheating, and we parameterize the
	process using a few key quantities: the duration of reheating, defined
	as
	$N_{\text{rh}} \equiv \ln \left( a_{\text{rh}} / a_{\text{end}}
	\right)$, where $a_{\text{end}}$ and $a_{\text{rh}}$ are the scale
	factors at the end of inflation and at the end of reheating,
	respectively; the effective equation-of-state (EoS) parameter during
	reheating, $\omega$, such that the total energy density evolves as
	$\rho \propto a^{-3(1+\omega)}$; and the reheating temperature,
	$T_{\text{rh}}$, defined as the temperature of the thermal bath at the
	end of reheating. We define the end of reheating via
	$\rho (a_{\text{end}}) =2 \rho_R (a_{\text{end}})$, where
	$\rho_R = g_\star \pi^2 T^4 / 30$ denotes the radiation energy density
	of the thermal bath with a temperature $T$. Here, $g_\star$
	corresponds to the effective degrees of freedom contributing to the
	energy density in radiation $\rho_R$.
	
	Depending on these parameters, the expansion history during reheating
	can vary, altering the time at which the CMB pivot scale reenters the
	horizon. This allows one to establish a relationship between the
	inflationary dynamics and the reheating parameters \cite{Dai:2014jja,
		Cook:2015vqa, Drewes:2017fmn}. In particular, we have
	\cite{Cook:2015vqa, Becker:2023tvd}:
	\begin{align}\label{eq:Trh_inflation}
		T_{\text{rh}} = \left[ \left( \frac{43} {11\, g_{\star s}(T_{\text{rh}})}
		\right)^{\frac{1}{3}} \frac {a_0 T_0} {k_\star} H_\star
		e^{-N_\star} \left( \frac{45 V_{\text{end}} } {\pi^2 g_{\star}(T_{\text{rh}})}
		\right)^{-\frac{1}{3\left(1+\omega\right)}}
		\right]^{\frac{3\left(1+\omega\right)}{3 \omega-1}}\,,
	\end{align}
	for $\omega \neq 1/3$. Here, $a_0$ and $T_0 = 2.73~\text{K}$
	correspond to the scale factor and temperature at present,
	respectively; $k_\star$ is the CMB pivot scale as mentioned in Section
	\ref{sec:Starobinsky}; $H_\star$ denotes the Hubble parameter when
	$\phi = \phi_\star$, given by
	$H_\star = \pi M_{\rm Pl} \sqrt{r(\phi_\star) A_s(\phi_\star) /2}$;
	$g_{\star s}$ is the effective degrees of freedom contributing to the
	total entropy densities; and $V_{\text{end}}$ is the potential energy
	at the end of inflation, i.e., when $\phi = \phi_{\text{end}}$
	(cf. Eq.~\eqref{eq:phiend}).
	
	Several remarks are in order regarding Eq.\eqref{eq:Trh_inflation}.
	We first note that the quantities with a $\star$ as subscript are
	defined as referring to the incident when the perturbation $k_\star$
	crossed out of the Hubble horizon during inflation, i.e.
	$k_\star = a_\star H_\star$; hence
	$a_0 H_\star/ k_\star = a_0 / a_\star$. For convenience we set
	$a_0 = 1$, which does not change the ratio $a_0/a_\star$. Moreover, we
	assume that only known particles in the SM contribute to $g_\star$ and
	$g_{\star s}$, in which case the effective number of relativistic
	degrees of freedom remains approximately constant at $106.75$ for
	temperatures above the top quark mass. At lower temperatures, however,
	$g_\star$ and $g_{\star s}$ vary; for instance, both are around
	$10.75$ when the reheating temperature is near the MeV scale. In our
	analysis, we have incorporated the temperature dependence of $g_\star$
	and $g_{\star s}$ \cite{Drees:2015exa}. We remind the reader that the
	inflationary predictions impacts $T_{\text{rh}}$ via the dependence on
	$H_\star$ and $N_\star$. Finally, we note that for the case with
	$\omega = 1/3$, the distinction between the reheating stage and the
	radiation-dominated phase becomes ambiguous, in which case a simple
	correlation between inflationary predictions and reheating parameters
	cannot be derived~\cite{Cook:2015vqa}. Indeed, for the special case
	with $\omega = 1/3$, one obtains the relation
	$61.5 \simeq N_\star + \ln\left( \frac{V^{1/4}_{\text{end}}}{H\star}
	\right)$, assuming $g_\star = g_{\star s} = 106.75$. This relation is
	independent of the reheating parameters and predicts a value of $n_s$
	for a given inflationary model. In particular, we find that it yields
	$n_s \simeq 0.965$ for the Starobinsky inflation model under
	consideration.
	
	\begin{figure}[ht]
		\def\sepf{0.8}
		\centering
		\includegraphics[scale=\sepf]{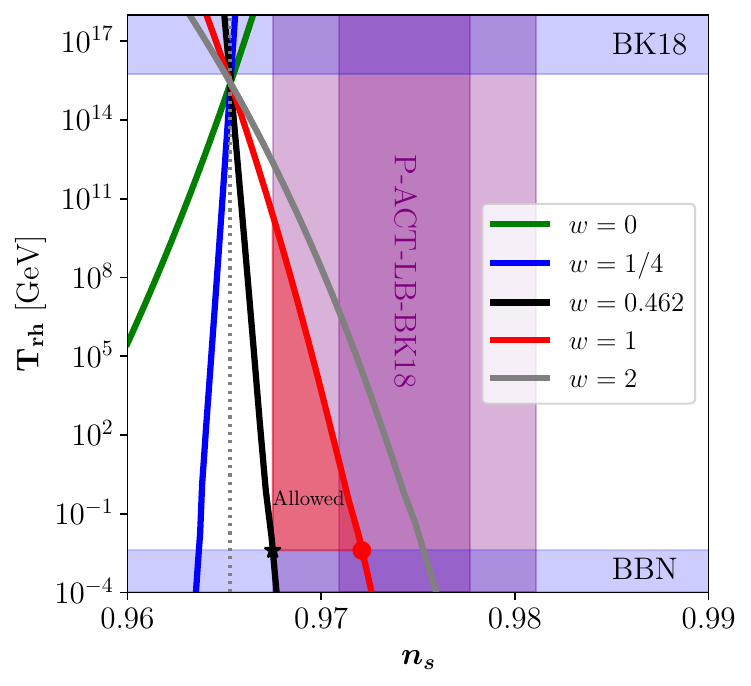}
		\caption{ Reheating temperature $T_{\text{rh}}$ as a function of $n_s$
			for different choices of the EoS parameter $\omega$ within the
			framework of Starobinsky inflation framework.  The vertical dotted
			gray line corresponds to $\omega =1/3$.}
		\label{fig:Trh_ns}
	\end{figure} 
	
	In Fig.~\ref{fig:Trh_ns}, we show the reheating temperature
	$T_{\text{rh}}$ as a function of $n_s$ and the EoS parameter $\omega$,
	together with the recent combined constraints on $n_s$ with ACT. The
	$1\sigma$ and $2 \sigma$ region are indicated by the purple
	band. Additionally, we depict the constraints on the reheating
	temperature, including $T_{\text{rh}} \geq 4~\text{MeV}$ from Big Bang
	Nucleosynthesis (BBN) \cite{Kawasaki:2000en, Hannestad:2004px,
		DeBernardis:2008zz, deSalas:2015glj} and
	$T_{\text{rh}} \leq 5.5 \times 10^{15}~\text{GeV}$ from the upper
	bound on the inflationary scale as measured by BICEP/KecK 2018
	\cite{BICEP:2021xfz}, shown in light blue.
	\begin{table}[h]
		\centering
		\renewcommand{\arraystretch}{1.5} 
		\setlength{\tabcolsep}{12pt} 
		\begin{tabular}{|c|c|}
			\hline
			$\omega$ & $\text{Maximum}\,\, T_{\text{rh}}~[\text{GeV}]$ \\
			\hline
			$0.462$ & $4 \times 10^{-3}$ \\
			$\frac{2}{3}$ & $1 \times 10^{7}$ \\
			$1$ & $2 \times 10^{10}$ \\
			$2$ & $2 \times 10^{12}$ \\
			\hline
		\end{tabular}
		\caption{Maximum value $T_{\text{rh}}$ as a function of $\omega$ being
			consistent with the $2\sigma$ constraint on $n_s$ from the combined
			results from ACT.}
		\label{tab1}
	\end{table}
	
	The solid green, blue, black, red, and gray lines correspond to
	$\omega = 0$, $\omega = 1/4$, $\omega = 0.462$, $\omega = 1$, and
	$\omega = 2$, respectively.  We note that the reheating temperature,
	$T_{\text{rh}}$, decreases (increases) with $n_s$ for $\omega > 1/3$
	($\omega < 1/3$), as shown in Fig.~\ref{fig:Trh_ns}. The special case
	$\omega = 1/3$ corresponds to a vertical line at $n_s \simeq 0.965$,
	independent of $T_{\text{rh}}$, as indicated by the vertical gray line
	in Fig.~\ref{fig:Trh_ns}. As $\omega$ approaches $1/3$, all curves are
	expected to converge toward the vertical gray line. In particular, a
	fixed point occurs at $T_{\text{rh}} \simeq 10^{15}\,\text{GeV}$ with
	$n_s \simeq 0.965$, where $d\, T_{\text{rh}} / d\, \omega \to 0$ for
	$\omega \neq 1/3$.
	
	For $\omega = 0$, we find that a reheating temperature
	$T_{\rm rh} \gtrsim 10^{20}$ GeV is required in order to be consistent
	with the $2\sigma$ value of $n_s$ from ACT; this is clearly
	unacceptable. $\omega = 0$ is often associated with perturbative
	reheating,\footnote{The numerical analysis of
		ref.~\cite{Ellis:2021kad}, which carefully models perturbative
		inflaton decays into radiation, finds a relation between
		$T_{\rm rh}$ and $n_s$ which is very close to our result for
		$\omega = 0$.} given that the potential \eqref{eq:V} is quadratic
	near the origin, $V(\phi) \simeq M^2 \phi^2 / 2 + {\cal
		O}(\phi^3)$. This also implies that the physical inflaton mass
	$m_\phi = M$. Since in a thermal bath the average particle energy
	$\langle E \rangle \sim 2.5 T$, perturbative reheating via inflaton
	decay requires $T_{\rm rh} < m_\phi/5$, unless reactions that {\em
		reduce} the number of particles (e.g., $3 \rightarrow 2$ reactions)
	are efficient, which seems highly implausible. For $\omega = 0$ and
	$T_{\rm rh} < 10^{13}$ GeV we find $n_s < 0.964$, which is three
	standard deviations below the ACT result.
	
	As $\omega$ increases, the curves shift from left to right, as seen in
	the green and blue curves. To satisfy the observational constraints,
	we find that $\omega \gtrsim 0.462$ is required. This is evident from
	the black solid line in Fig.~\ref{fig:Trh_ns}, which touches the
	crossing point between the lower bound on $T_{\text{rh}}$ from BBN and
	the $2\sigma$ boundary from ACT. As $\omega$ increases further, the
	overlapping region expands from a point (the black star) into a finite
	area. The red line ($\omega = 1$) crosses
	$T_{\text{rh}} = 4~\text{MeV}$ at $n_s \simeq 0.972$, corresponding to
	$N_\star \simeq 69$; see also the red dot in Fig.~\ref{fig:Trh_ns}. In
	this case, the allowed reheating parameter space for $\omega \leq 1$
	is represented by the red-shaded region.
	
	In Table \ref{tab1}, we show the maximum reheating temperature as a
	function of $\omega$. Assuming $\omega \leq 1$, the combined
	constraints for reheating parameters are found to be
	\begin{align} \label{eq:combined_constraints}
		0.462 \lesssim \omega\,; \quad 4~\text{MeV} \leq T_{\text{rh}}  \lesssim 2 \times 10^{10}~\text{GeV}\,,
	\end{align} 
	together with the inflationary parameters
	\begin{align}\label{eq:combined_ns}
		0.967 \lesssim n_s\lesssim 0.972\,;\quad  58 \lesssim N_\star \lesssim 69 \,,
	\end{align} 
	in order to be consistent with the recent combined results on $n_s$
	from ACT at $2\sigma$ level.  Similarly, we can derive more stringent
	constraints at the $1\sigma$ level, finding that
	$0.804 \lesssim \omega$ and
	$4~\text{MeV} \lesssim T_{\text{rh}} \lesssim 10~\text{GeV}$, with
	$0.971 \lesssim n_s \lesssim 0.972$ and
	$66 \lesssim N_\star \lesssim 69$. We note that for $\omega > 1$, the
	upper bounds on $T_{\text{rh}}$ and $N_\star$ can be larger. For
	example, for $\omega = 2$, the reheating temperature can reach
	$T_{\text{rh}} \simeq 10^{12}~\text{GeV}$ and $N_\star \simeq 78$ at
	the $2\sigma$ level of $n_s$. We note that a large EoS parameter,
	$\omega > 0.5$, may be realized if the dynamics after the end of
	Starobinsky inflation transition to those driven by a potential
	significantly steeper than quadratic. For instance, if the inflaton
	oscillates around a potential of the form\footnote{We
		note that a potential of this form is possible within the
		framework of $\alpha$-attractor scenarios, such as the
		$T$-model~\cite{Kallosh:2013yoa}.} $\phi^n$, with $n$ being a
	positive even integer, the EoS parameter is given by
	$\omega = (n - 2)/(n + 2)$ \cite{Turner:1983he}, implying that
	$\omega$ can approach unity for large $n$.  Constructing a viable
	theory within the current framework that supports such large values of
	$\omega$ may nevertheless present challenges.
	
\begin{figure}[ht]
		\def\sepf{0.8}
		\centering
		\includegraphics[scale=\sepf]{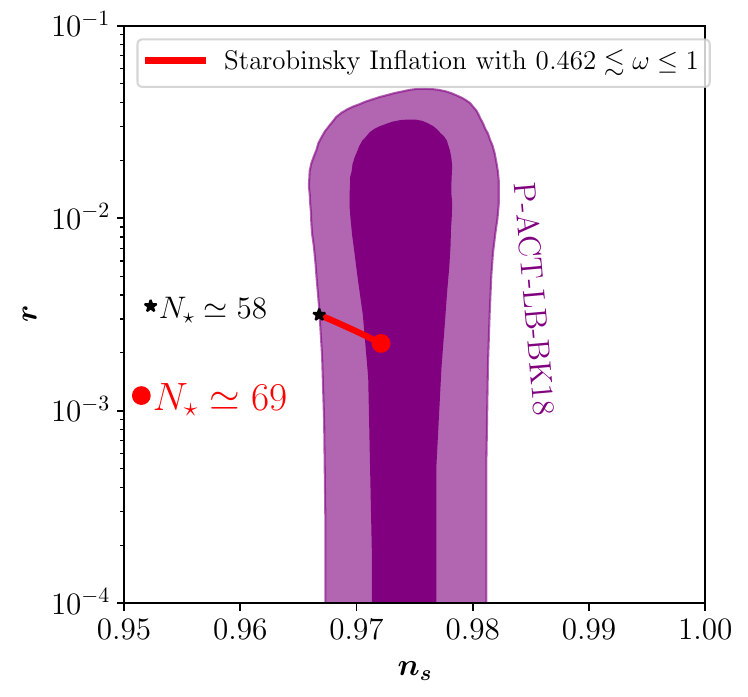}
		\caption{Inflationary predictions for the Starobinsky model with an
			effective post-inflationary EoS parameter
			$0.462 \lesssim \omega \leq 1$ and $T_\text{rh} \geq
			4~\text{MeV}$. The light purple region corresponds to the $2\sigma $
			allowed region of $n_s$ from ACT.}
		\label{fig:ns_r_combined}
	\end{figure}

	In Fig.~\ref{fig:ns_r_combined}, we depict the inflationary prediction
	(red solid line) consistent with the $2\sigma$ region of $n_s$ from
	ACT, considering $0.462 \lesssim \omega \leq 1$. The corresponding
	predictions for the tensor-to-scalar ratio are
	\begin{align}
		2 \times 10^{-3} \lesssim r \lesssim 3 \times 10^{-3}\,,
	\end{align}
	which can be tested by future precise cosmic microwave background
	(CMB) experiments. For instance, CORE~\cite{COrE:2011bfs},
	AliCPT~\cite{Li:2017drr}, LiteBIRD~\cite{Matsumura:2013aja}, and
	CMB-S4~\cite{Abazajian:2019eic} are expected to achieve sensitivity to
	$r \sim \mathcal{O}(10^{-3})$.
	
	\section{Conclusion}\label{sec:conclusion}
	
	In this work, we revisit the inflationary predictions of the
	Starobinsky model and investigate their compatibility with
	post-inflationary reheating dynamics, ensuring consistency with recent
	combined measurements of the spectral index from the Atacama Cosmology
	Telescope (ACT).
	
	We present a refined approximation in Eq.~\eqref{eq:ns_iteration},
	showing that when evaluated using these more accurate expressions, the
	predictions of the Starobinsky model remain consistent with the recent
	combined results from ACT at the $2\sigma$ level, as illustrated in
	Fig.~\ref{fig:ns_r}. Relying on the commonly used approximations
	provided in Eq.~\eqref{eq:approximation} underestimates $n_s$, leading
	to an observation that the Starobinsky model with $N_\star \simeq 60$
	lies at the edge of the $2\sigma$ constraints.
	
	We also investigate the constraints on the reheating parameters. To
	this end, we parameterize reheating using an effective equation of
	state (EoS) parameter $\omega$, the reheating temperature
	$T_{\text{rh}}$, and the duration of reheating. The main results are
	presented in Fig.~\ref{fig:Trh_ns} and Table~\ref{tab1}. Purely
	perturbative reheating, which leads to $\omega \simeq 0$ and
	$T_{\rm rh} \lesssim 10^{13}$ GeV, is disfavored at $\gtrsim3
	\sigma$. We find that $\omega \gtrsim 0.462$ is needed in order to be
	consistent with Big Bang Nucleosynthesis (BBN) as well as the
	$2\sigma$ region of the combined ACT results for $n_s$. For
	$\omega \leq 1$, we derive an upper bound on the reheating temperature
	of approximately $T_{\text{rh}} \lesssim 2 \times
	10^{10}~\text{GeV}$. These results are summarized in
	Eq.~\eqref{eq:combined_constraints}. The combined inflationary
	predictions consistent with BBN as well as the $2\sigma$ region of
	$n_s$ from ACT results are depicted in Fig.~\ref{fig:ns_r_combined}
	and can be tested by future cosmic microwave background (CMB)
	experiments.
	
	In summary, we have provided refined approximations for the
	inflationary predictions of the Starobinsky model and derived
	constraints on the reheating parameters based on the latest combined
	experimental results on $n_s$ with ACT.

	\section*{Acknowledgments}
	YX thanks Carlos Tamarit and Felix Yu for discussions. YX acknowledges
	the support from the Cluster of Excellence ``Precision Physics,
	Fundamental Interactions, and Structure of Matter'' (PRISMA$^+$ EXC
	2118/1) funded by the Deutsche Forschungsgemeinschaft (DFG, German
	Research Foundation) within the German Excellence Strategy (Project
	No. 390831469).
	
	\bibliographystyle{JHEP} \bibliography{biblio}
\end{document}